\documentclass[12pt,3p,preprint,sort&compress]{elsarticle}
\usepackage[colorlinks=true]{hyperref}

\usepackage{amsmath}
\usepackage{amssymb}
\usepackage{mathtools}
\usepackage[utf8]{inputenc}

\renewcommand{\l}{\left(}
\renewcommand{\r}{\right)}

\newcommand{\be}{\begin{equation}}
\newcommand{\ee}{\end{equation}}
\newcommand{\ba}{\begin{align}}
\newcommand{\ea}{\end{align}}
\newcommand{\bg}{\begin{gather}}
\newcommand{\eg}{\end{gather}}
\newcommand{\bseq}{\begin{subequations}}
\newcommand{\eseq}{\end{subequations}}

\newcommand{\half}{\frac{1}{2}}

\begin{document}
\begin{flushright}
	INR-TH-2025-014
\end{flushright}

\title{Decays and annihilation of galactic dark matter: \\determine $D$-, $J_s$-, $J_p$- and $J_d$-factors with dark matter profiles inferred from GravSphere fit to stellar observations} 

\author[ug]{Fedor Bezrukov}
\author[inr,mpti]{Dmitry Gorbunov}
\ead{gorby@ms2.inr.ac.ru}
\author[inr,msu]{Ekaterina Koreshkova}

\address[ug]{Department of Microbiology and Molecular Medicine, University of Geneva,
    1211 Geneva, Switzerland}
\address[inr]{Institute for Nuclear Research of Russian Academy of Sciences, 117312 Moscow, Russia}
\address[mpti]{Moscow Institute of Physics and Technology, 141700 Dolgoprudny, Russia} 
\address[msu]{Department of Particle Physics and Cosmology, Physics Faculty, M.V. Lomonosov Moscow State University,
    Vorobjevy Gory, 119991 Moscow, Russia}

\begin{abstract}
    Dark matter mass density profiles and velocity distributions for a set of dwarf spheroidal galaxies (dSphs) have recently been obtained \cite{Bezrukov:2024qwr} by performing a multi-parametric fit to the stellar observations with the help of the GravSphere which solves the Jeans equation. We use these results to calculate the geometrical factors for estimation of the fluxes of cosmic rays expected from decay ($D$-factor) and annihilation ($J_s$-, $J_p$- and $J_d$-factors for $s$-, $p$- and $d$- wave processes) of dark matter particles in galaxies. The general novelty is the account for a possible anisotropy in velocities of dark matter particles. On the basis of this analysis we present empirical scaling approximations to these factors as functions of typical observables:  distance to the galaxy $d$, it's half-radius $r_h$ and line-of-sight stellar velocity dispersion $\sigma_{LOS}$. They can be applied to any galaxy, and for $D$- and $J_s$-factors we refine the estimates of Ref.\,\cite{Pace:2018tin}: the shifts in the central values remain within 1-2$\sigma$ error bars. In particular, the simplest approximations for the factors calculated within a fixed viewing angle (with exponents fixed from geometrical considerations) are:  
    
    $\log_{10}\l D(\theta_{max})/\text{GeV}^{-1}\text{cm}^2\r=\l 16.63\pm0.01 \r \l \sigma_{LOS}/5\,\text{km/s}\r^2\l d/100\,\text{kpc}\r^{-2}\l r_h/100\,\text{pc}\r$, 

$\log_{10} \l J_s(0.5^\circ)/\text{GeV}^2\text{cm}^5\r= \l 17.92\pm 0.04\r\l \sigma_{LOS}/5\,\text{km/s}\r^4\l d/100\,\text{kpc}\r^{-2}\l r_h/100\,\text{pc}\r^{-1}$,

$\log_{10} \l J_p(0.5^\circ)/\text{GeV}^2\text{cm}^5\r= \l 15.59\pm 0.08\r\l \sigma_{LOS}/5\,\text{km/s}\r^6\l d/100\,\text{kpc}\r^{-2}\l r_h/100\,\text{pc}\r^{-1}$,

$\log_{10} \l J_d(0.5^\circ)/\text{GeV}^2\text{cm}^5\r=\l 13.68\pm0.11\r \l \sigma_{LOS}/5\,\text{km/s}\r^8\l d/100\,\text{kpc}\r^{-2}\l r_h/100\,\text{pc}\r^{-1}$.

\end{abstract}
\date{}

\maketitle


{\bf 1.} Searches for possible annihilation or decays of dark matter (DM) particles in cosmic structures, galaxies and galaxy clusters, are important tasks for many astronomical instruments. Note, that it is not guaranteed, that any DM is really present in the galaxies, as potentially some (unfortunately not yet formulated in a consistent way) modification of gravity could be responsible for the observed phenomena usually attributed to the DM. At the same time, many DM models predict such processes, and negative results of searches for the expected signals in photon, neutrino and cosmic ray spectra constrain the model parameter spaces, see e.g.\,\cite{Krivonos:2024yvm,Kachelriess:2018rty,Cuoco:2017iax}. 

These DM models are different in many aspects, yet the common feature is a coupling of dark matter sector to the Standard Model particles which ensures the production of the latter in decays and  annihilation of the former. The mass of DM particles and the type of coupling vary from model to model and so does the signal that could be observed at the Earth. The strength of the signal depends on the DM decay rate $\Gamma_{\text{DM}}$ or the DM annihilation cross section $\sigma_{\text{DM}}(v_{\text{DM}})$, where $v_{DM}$ is the average velocity of the DM particles at annihilation. The decay rate and cross section are generally unique for the DM models and the signals under discussion would (in principle) allow one to distinguish the DM model; for reviews on indirect DM searches see, e.g.,\,\cite{Ibarra:2013cra,Gaskins:2016cha,PerezdelosHeros:2020qyt}. 

The signal from a given part of the galaxy also depends on the local density of DM particles $n_{DM}$. The signal from a given direction depends on the dark matter density along the line of sight. The total signal flux from a far galaxy is fixed by the space integral involving the DM density profile. The DM profile in a given galaxy is the same for all the DM models as far as the DM self-interaction and couplings to the visible matter play no role in the galaxy formation. This is the case for the Weakly Interacting Massive particles, sterile neutrinos, WIMPzillas and many other models, where the galaxy formation is the gravity solo performance. In all these models the signal flux, expected from annihilation of decay of the DM particles, is proportional to specific geometrical integral factors which are entirely determined by the DM density and velocity profiles in the galaxy. Both quantities, especially the latter, cannot be directly extracted from observations.

{\bf 2.} In a given galaxy the DM profile $\rho(r)$ (the matter density as a function of the distance to the galaxy centre $r$) and its velocity distribution $f(v,r)$ can be inferred by fitting observations of the positions and line-of-sight velocities of galaxy stars. In this work we use the results of Ref.\,\cite{Bezrukov:2024qwr}, where 20 dwarf spheroidal galaxies have been studied applying the fitting procedure performed with the help of GravSphere\,\cite{GravSphere_ref}.\footnote{\url{https://github.com/justinread/gravsphere}}. 

\begin{figure}
    \centerline{
    \includegraphics[width=0.33\linewidth]{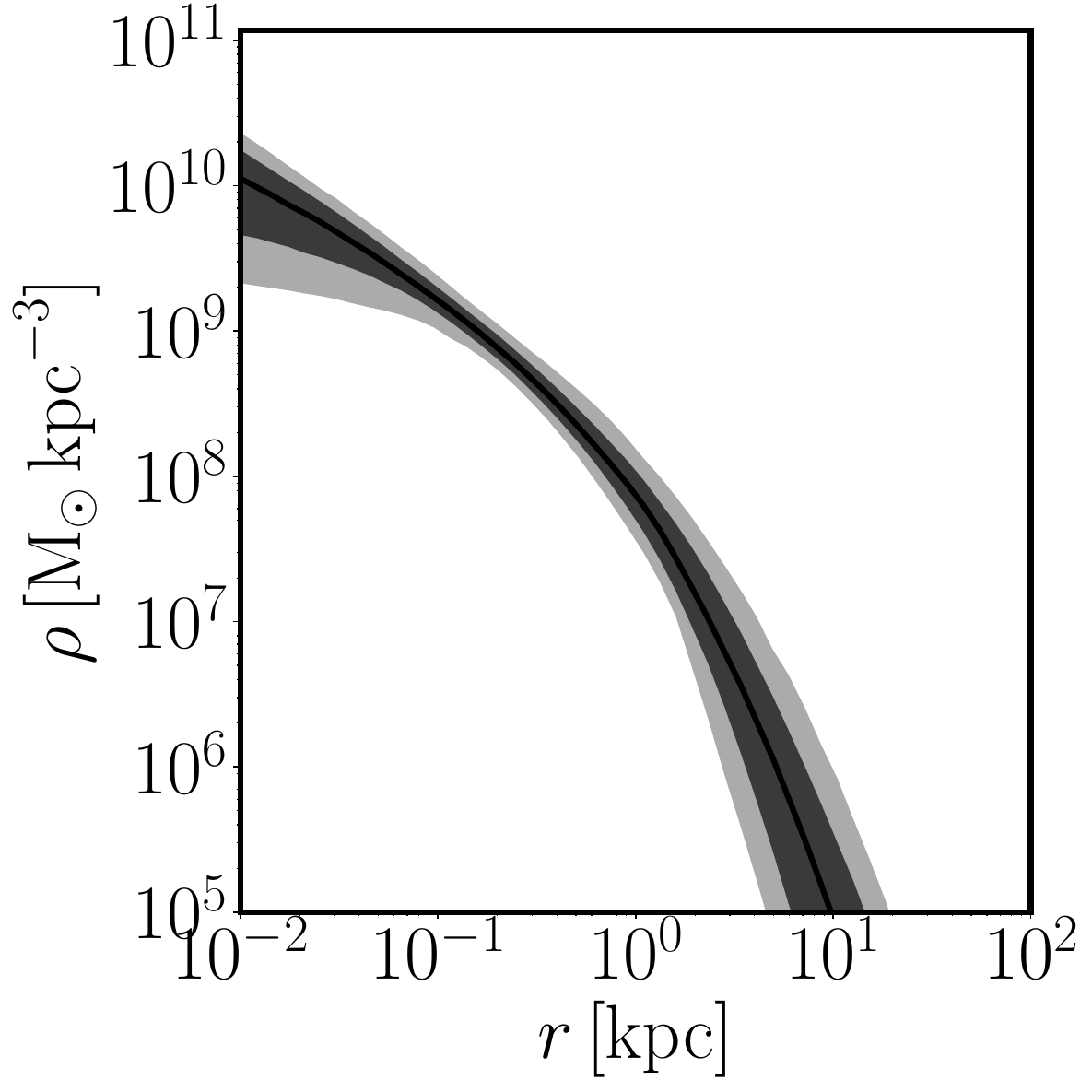}
    \includegraphics[width=0.33\linewidth]{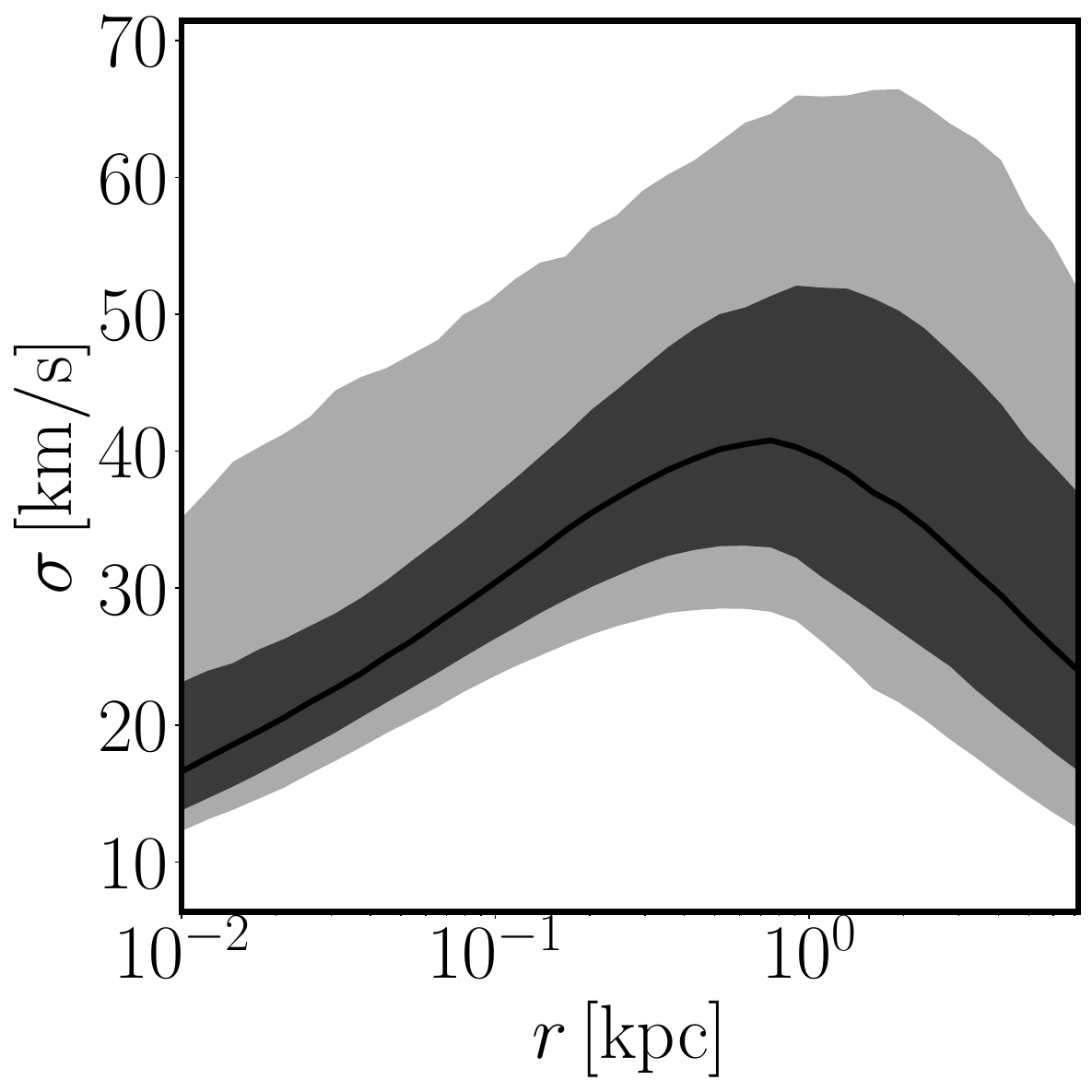}
    \includegraphics[width=0.33\linewidth]{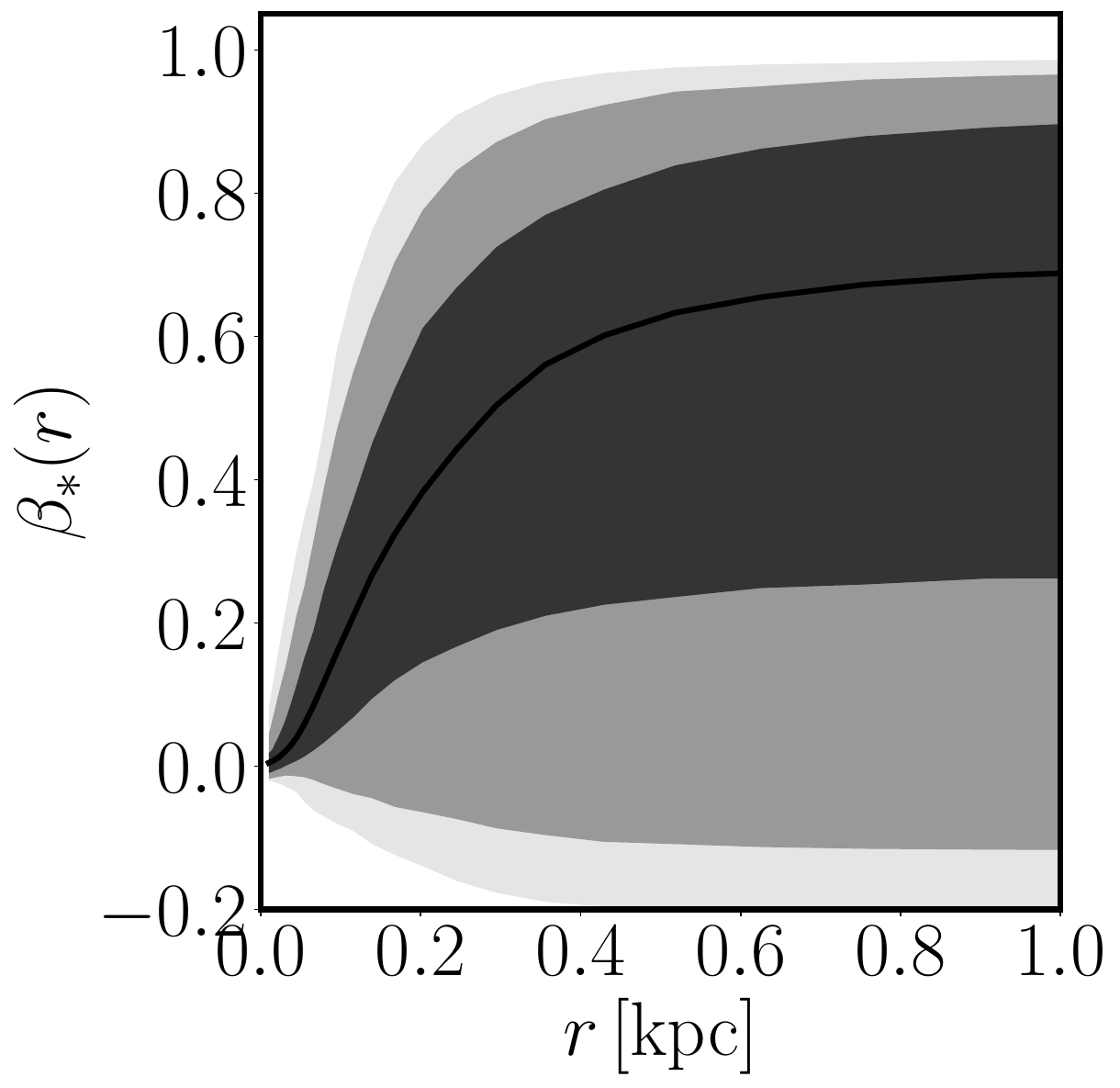}
    }
    \centerline{
    \includegraphics[width=0.33\linewidth]{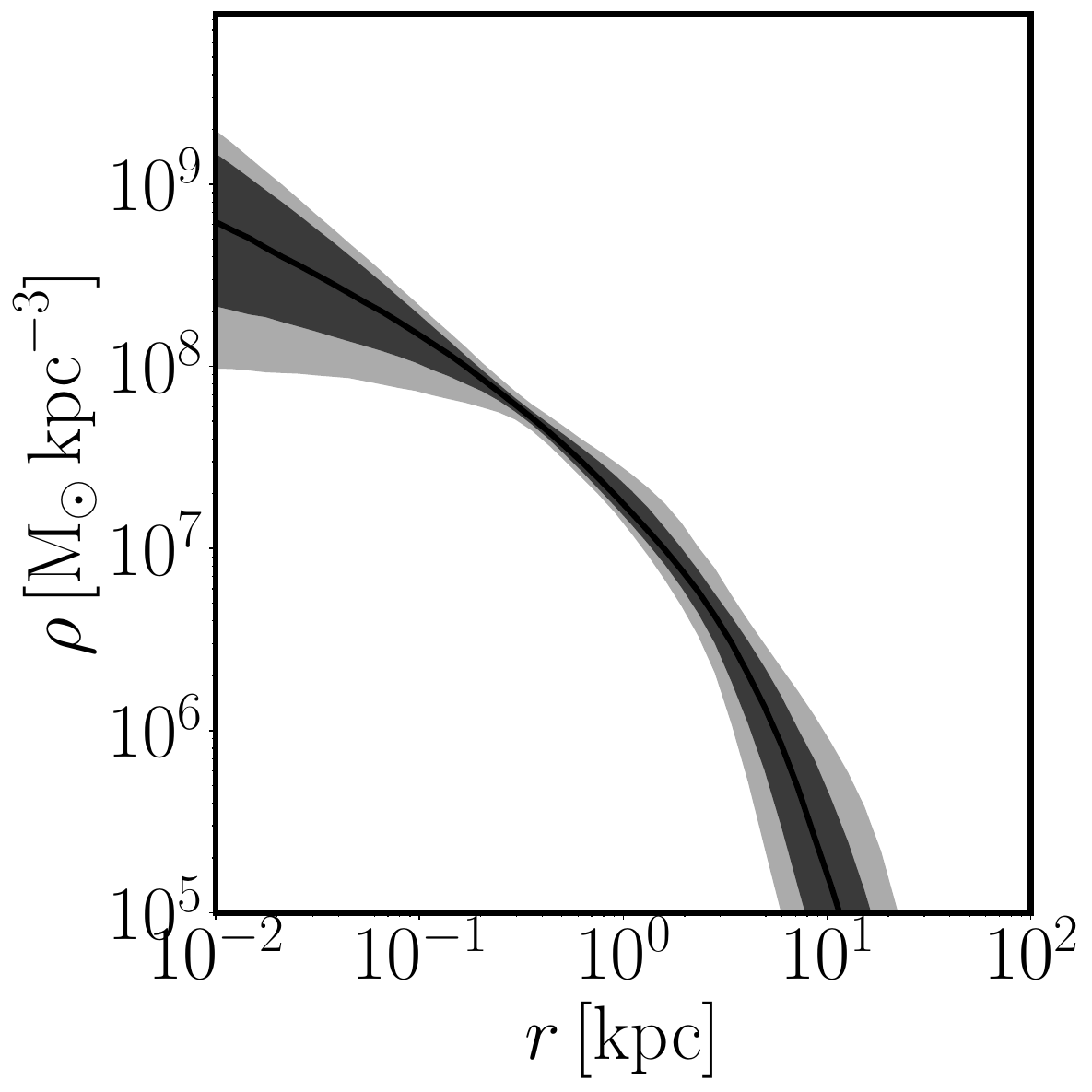}
    \includegraphics[width=0.33\linewidth]{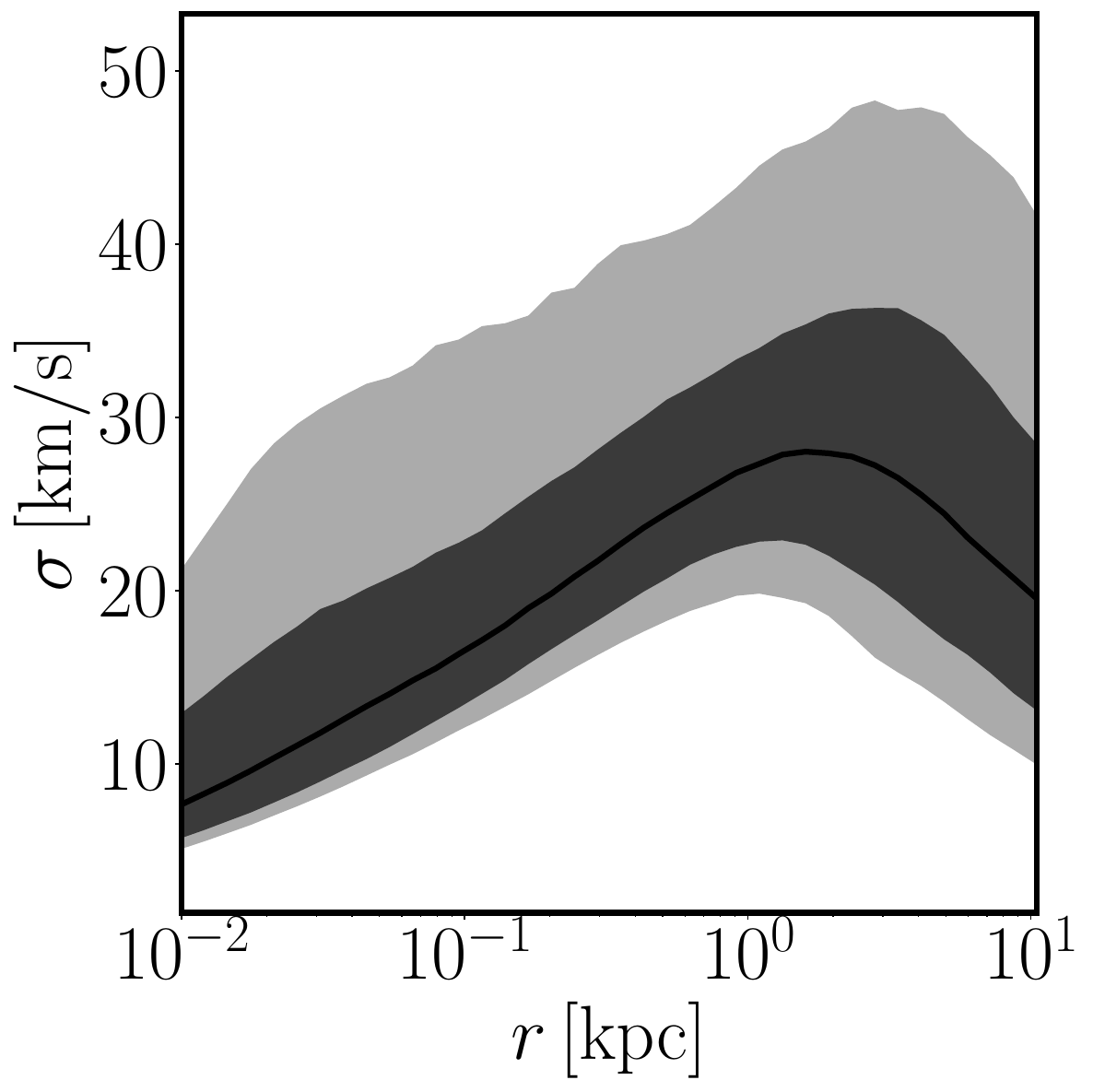}
    \includegraphics[width=0.33\linewidth]{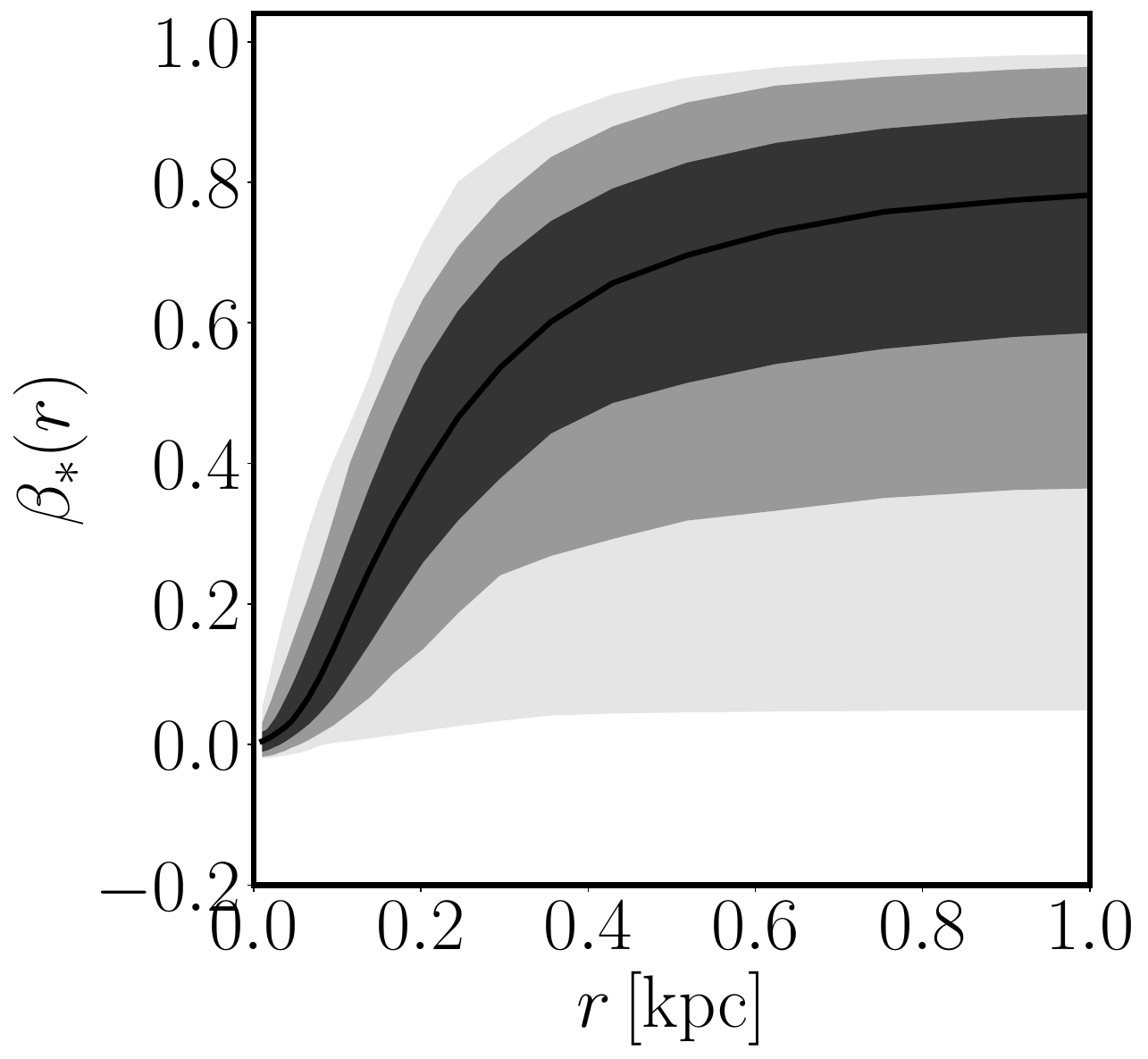}
    }
    \centerline{
    \includegraphics[width=0.33\linewidth]{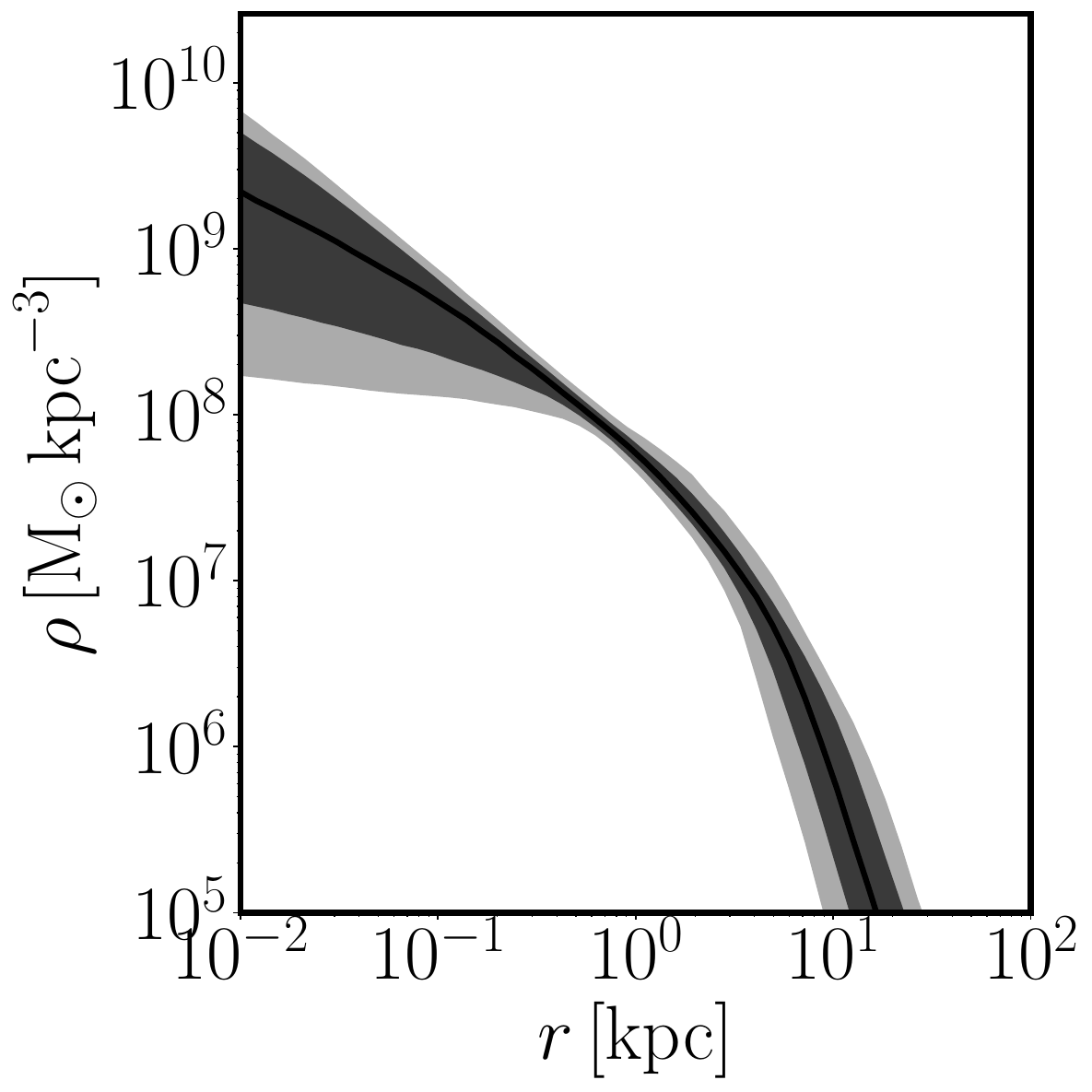}
    \includegraphics[width=0.33\linewidth]{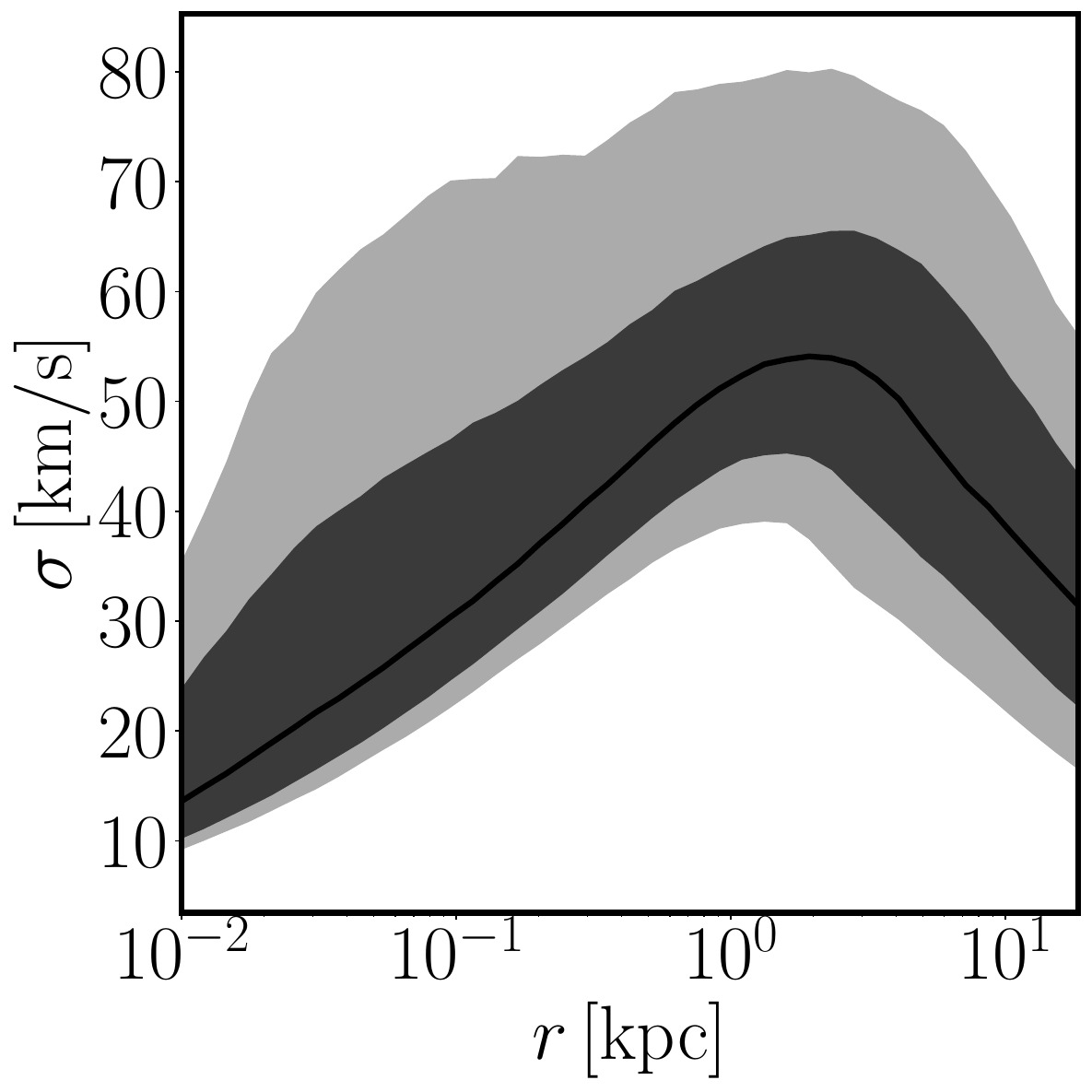}
    \includegraphics[width=0.33\linewidth]{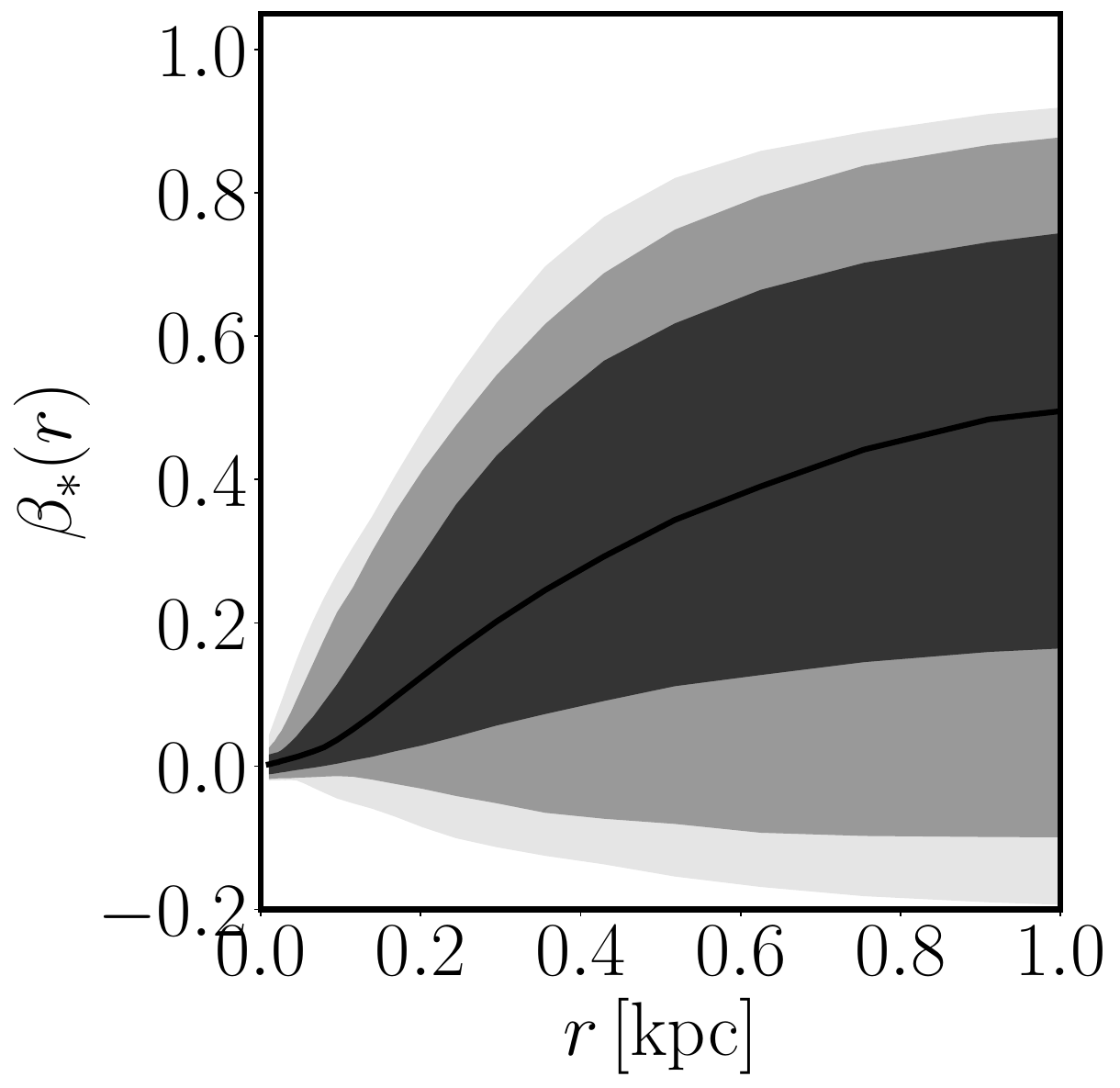}
    }
    \caption{The obtained DM profiles (left panels), DM velocity dispersions (central panels) and {\it stellar} velocity anisotropies (right panels) for DM populations of the Bootes Dwarf Spheroidal Galaxy (top panels),  Carina dSph (middle panels) and Sextans dSph (bottom panels). The light (dark) gray colors outline the 1(2)\,$\sigma$ regions, see Ref.\,\cite{Bezrukov:2024qwr} for details and \cite{Koreshk_github}, where similar plots for other galaxies from Tab.\,\ref{tab:results} are presented.}
    \label{fig:profiles}
\end{figure}

The set of galaxies, see Tab.\,\ref{tab:results}, has been chosen by requiring a sufficient amount of data to be used by the algorithm which solves the Jeans equation for stars with a multi-parametric DM profile. 
The velocity of stars is fitted with the Maxwell distribution,  
\begin{equation}
\label{maxwell}
f(v_r,v_\tau,r)=\frac{1}{\l 2\pi\r^{3/2}\sigma_r\sigma^2_\tau}\exp{\l-\half \l\frac{v_r^2}{\sigma_r^2} +\frac{v_\tau^2}{\sigma_\tau^2}\r\r }\,,
\end{equation}
where two parameters characterize the radial $\sigma_r(r)$ and tangential $\sigma_\tau(r)$ stellar velocity dispersions as functions of the radius $r$. The stellar velocity dispersions were supposed to be related parametrically as follows,  
\begin{equation}
\label{asymmetry}
\sigma_\tau^2=(1-\beta(r))\sigma_r^2\,,\;\;\;\;\;\beta(r)=\beta_0+\frac{\beta_\infty-\beta_0}{1+\l r/r_0\r^n}\,.
\end{equation}
The program generates the Monte Carlo chains of the specifically parametrized profiles $\rho(r)$ and stellar velocities (fitting parameters $\beta_0$, $\beta_\infty$, $r_0$ and $n$) that fit the data and obey the Jeans equation, for details see Refs.\,\cite{Read_2018,Collins_2021}. The procedure yields a distribution of profiles according to the goodness of fit, which allows one to infer the best fit profile and its variance; see the left panels of Fig.\,\ref{fig:profiles} 
for examples\footnote{Profiles for other galaxies can be found following the link \cite{Koreshk_github}.}. 

To estimate the velocity distributions of DM particles, Ref.\,\cite{Bezrukov:2024qwr} accepted the Maxwell form similar to \eqref{maxwell} and similar to \eqref{asymmetry} relation between the radial and tangentialDM velocity variances, that characterize the DM velocity asymmetry. The GravSphere code has been modified by adding the solution of the Jeans equation for the DM particles. The parameters of DM velocity distribution for each profile, determined by means of the Monte Carlo simulations, were randomly chosen following the flat distributions  
$\beta_\infty\in[0;0.56]$, $r_0\in[0;3.10\,\text{kpc}]$, $n\in[0.73;1.36]$ while keeping $\beta_0=0$. The examples of the obtained DM velocity distributions\,
\footnote{Velocity distributions of DM particles for other galaxies can be found following the link \cite{Koreshk_github}.} 
are shown on the right panels of Fig.\,\ref{fig:profiles}. 

{\bf 3.} DM particles, decaying and annihilating into the Standard Model particles, contribute to the flux of cosmic particles in a given solid angle $d\Omega$, and the signal is proportional to the integral along the line-of-sight $dl$. Namely, the signal of DM decay is proportional to $D$-factor defined as 
\begin{equation}
    \label{D-factor}
    D\equiv \int d\Omega \int dl \,\rho(r) \,.
\end{equation}

The local signal from the DM annihilation is proportional to the squared number density of DM particles and the annihilation cross section $\sigma_{ann}$ multiplied by the relative velocity $v_{rel}$ of DM particles. The galactic DM particles are non-relativistic, and so the interesting product may be cast into series of degrees of the squared DM relative velocity 
\[
\sigma_{ann}v_{rel}=\sigma_0+\sigma_1 v_{rel}^2+\sigma_2v_{rel}^4+\dots
\]
In quantum mechanics each term in the expansion corresponds to the  contribution of a particular wave of the final state, formed by the outgoing particles.  Since the annihilating particles are non-relativistic, the contributions of the higher terms are naturally suppressed, so the contribution of $s$-wave, i.e. $\sigma_0$, dominates. However, the microscopic physics, responsible for the annihilation, may result in a numerical suppression of this term, so the annihilation proceeds mostly in $p$-wave, or even $d$-wave. According to these options, the corresponding $J_s$-, $J_p$- and $J_d$-factors are defined as the following integrals over the line-of-sight: 
\begin{align}
\label{Js}
    J_s&\equiv \int d\Omega \int dl \,\rho^2(r) \,,\\
     J_p&\equiv \int d\Omega \int dl \,\rho^2(r)\int d^3v \int d^3u f(u,r)f(v,r)v_{rel}^2 \,,\\
J_d&\equiv \int d\Omega \int dl \,\rho^2(r)\int d^3v \int d^3u f(u,r)f(v,r)v_{rel}^4 \,,
\end{align}
with relative 3-velocity of the annihilating particles $v_{rel}=v-u$. 

The first factor $J_s$ can be straightforwardly calculated for all 20 galaxies by integrating the squared DM profiles obtained in Ref.\,\cite{Bezrukov:2024qwr}. To estimate the other two factors, one must integrate over the DM velocities. The squared relative velocity can be presented via radial and tangential projections of the velocities of two annihilating particles and the angles between the tangential components, namely: 
\[
v_{rel}^2=(v_r-u_r)^2+v_\tau^2+u_\tau^2-2 v_\tau u_\tau \cos{(\phi_v-\phi_u)}\,.
\]
The integration over the two angles cancels the contribution of very last term, so we get
\[
\int d\phi_v\int d\phi_u v_{rel}^2= 4\pi^2 \l (v_r-u_r)^2+v_\tau^2+u_\tau^2\r\,. 
\]
Then, substituting the Maxwell distribution \eqref{maxwell} for each of the annihilating particles, we obtain 
\begin{equation}
    \label{}
    \begin{split}
& \int d^3v \int d^3u f(u,r)f(v,r)v_{rel}^2 \\ &= \int\frac{dv_r du_r dv_\tau^2 du_\tau^2}{8\pi \sigma_r^2\sigma_\tau^4} \l (v_r-u_r)^2+v_\tau^2+u_\tau^2\r \exp{\l -\frac{v_r^2+u_r^2}{2\sigma_r^2}- 
\frac{v_\tau^2+u_\tau^2}{2\sigma_\tau^2}\r} \\
&=\int \frac{dv_r du_r}{2\pi \sigma_r^2} \l (v_r-u_r)^2+
4\sigma^2_\tau\r\exp{\l -\frac{v_r^2+u_r^2}{2\sigma_r^2}\r}=2\sigma_r^2+4\sigma_\tau^2\,.
\end{split}
\end{equation}
Consequently for the $p$-wave factor we get in terms of DM profile, radial velocity dispersion and asymmetry 
\begin{equation}
\label{Jp}
J_p=\int d\Omega \int dl \rho^2(r)\,2\sigma_r^2(r)\l 3-2\beta(r)\r\,. 
\end{equation}
Similarly, for the $d$-wave factor one finds 
\begin{equation}
\label{Jd}
J_d=\int d\Omega \int dl \rho^2(r)4\sigma_r^4(r)\l 15-20\beta(r)+8\beta^2(r)\r\,.
\end{equation}

We calculate the $D$-factor and $J$-factors by  substituting in formulas \eqref{D-factor}, \eqref{Js}, \eqref{Jp}, \eqref{Jd} the DM profiles $\rho(r)$, velocity 
radial dispersions $\sigma_r^2(r)$ and asymmetry factors $\beta(r)$ obtained in Ref.\,\cite{Bezrukov:2024qwr} and presented in \cite{Koreshk_github} for each galaxy. According to Ref.\,\cite{Ackermann_2011} the angular size of a dwarf galaxy in the indirect searches for the DM is about $1^\circ$. Hence in the above integration the solid angle is taken to be $\mathrm{d}\Omega=2\pi\sin\theta\mathrm{d}\theta$ with $\theta\in[0;0.5^\circ]$. The integration over the line of sight is performed within the range $l\in[l_-;l_+]$, where $l_\pm=d\cos\theta\pm\sqrt{(250\,\text{kpc})^2-(d\sin\theta)^2}$ and $d$ is a distance to the galaxy. 
In the GravSphere the radius goes up to $250$\,kpc, but as far as all the factors are we are calculating are saturated at the size of the order of the visible size of the galaxy, or the half-light radius $r_h$, which are typically at least an order of magnitude smaller than 250\,kpc, this is not a limitation.

The results of the integral evaluation are presented in Tab.\,\ref{tab:results}. 
\begin{table}[!htb]
\centering    
\resizebox{\columnwidth}{!}{
    \begin{tabular}{|l|c|c|c|c|c|c|}
    
    \hline
        Name in SIMBAD database& $d/$\text{kpc}& $\log_{10}\frac{J(0.5^\circ)}{ \text{GeV}^2\text{cm}^{-5}}$ & $\log_{10}\frac{J_p(0.5^\circ)}{ \text{GeV}^{2}\text{cm}^{-5}}$ & 
        $\log_{10}\frac{J_d(0.5^\circ)}{\text{GeV}^{2}\text{cm}^{-5}}$ &
        $\log_{10}\frac{D(0.5^\circ)}{\text{GeV}\;\text{cm}^{-2}}$ &  $\log_{10}\frac{D(\theta_{max})}{\text{GeV}\;\text{cm}^{-2}}$ \\ \hline  \hline
        
        Andromeda V & $810\pm45$ & $16.76^{+0.11}_{-0.10}$ & $ 14.90 ^{+ 0.30 } _{- 0.23 }$ & $ 13.46 ^{+ 0.40 } _{- 0.35 }$
        & $ 17.21 ^{+ 0.24 } _{- 0.20 }$ & $16.11^{+0.05}_{-0.05}$
        \\ \hline
        
        Aquarius Dwarf & $940\pm38$ & $16.75^{+0.19}_{-0.17}$ & $ 15.18 ^{+ 0.39 } _{- 0.43 }$ & $ 13.82 ^{+ 0.53 } _{- 0.58 }$
        & $ 17.19 ^{+ 0.34 } _{- 0.32 }$ & $15.81^{+0.07}_{-0.07}$
        \\ \hline
        
        Bootes Dwarf Spheroidal Galaxy & $66\pm3$ & $20.22^{+0.18}_{-0.16}$ & $ 18.83 ^{+ 0.47 } _{- 0.41 }$ & $ 18.07 ^{+ 0.62 } _{- 0.54 }$
        & $ 19.14 ^{+ 0.14 } _{- 0.11 }$ & $18.26^{+0.09}_{-0.08}$
        \\ \hline
        
        Carina dSph & $105\pm6$ & $18.32^{+0.11}_{-0.08}$ & $ 16.92 ^{+ 0.25 } _{- 0.24 }$ & $ 15.77 ^{+ 0.40 } _{- 0.36 }$
        & $ 18.44 ^{+ 0.09 } _{- 0.07 }$ & $17.39^{+0.04}_{-0.04}$
        \\ \hline
        
        Cetus Dwarf Galaxy & $790\pm40$ & $16.25^{+0.22}_{-0.18}$ & $ 13.70 ^{+ 0.37 } _{- 0.38 }$ & $ 11.81 ^{+ 0.54 } _{- 0.46 }$
        & $ 16.67 ^{+ 0.28 } _{- 0.22 }$ & $15.91^{+0.06}_{-0.08}$
        \\ \hline
        
        Coma Dwarf Galaxy & $44\pm4$ & $19.08^{+0.34}_{-0.52}$ & $ 16.22 ^{+ 0.30 } _{- 0.32 }$ & $ 14.48 ^{+ 0.38 } _{- 0.40 }$
        & $ 18.42 ^{+ 0.10 } _{- 0.13 }$ & $17.93^{+0.14}_{-0.22}$
        \\ \hline
        
        CVn I dSph & $218\pm10$ & $17.43^{+0.17}_{-0.15}$ & $ 15.74 ^{+ 0.27 } _{- 0.22 }$ & $ 14.30 ^{+ 0.37 } _{- 0.34 }$
        & $ 17.98 ^{+ 0.10 } _{- 0.09 }$ & $17.07^{+0.07}_{-0.07}$
        \\ \hline
        
        Dra dSph & $76\pm5$ & $19.56^{+0.14}_{-0.14}$ & $ 18.51 ^{+ 0.35 } _{- 0.32 }$ & $ 17.70 ^{+ 0.49 } _{- 0.43 }$
        & $ 18.97 ^{+ 0.10 } _{- 0.10 }$ & $17.95^{+0.06}_{-0.06}$
        \\ \hline
        
        Fornax Dwarf Spheroidal & $147\pm12$ & $18.09^{+0.03}_{-0.03}$ & $ 16.63 ^{+ 0.08 } _{- 0.07 }$ & $ 15.41 ^{+ 0.14 } _{- 0.13 }$
        & $ 18.33 ^{+ 0.02 } _{- 0.02 }$ & $17.82^{+0.01}_{-0.01}$
        \\ \hline
        
        Hercules Dwarf Galaxy & $132\pm6$ & $17.10^{+0.36}_{-0.46}$ & $ 13.89 ^{+ 1.03 } _{- 0.72 }$ & $ 11.66 ^{+ 1.34 } _{- 1.04 }$ 
        & $ 17.41 ^{+ 0.34 } _{- 0.26 }$ & $16.80^{+0.19}_{-0.19}$
        \\ \hline
        
        Leo A & $800\pm100$ & $16.53^{+0.21}_{-0.16}$ & $ 14.36 ^{+ 0.37 } _{- 0.41 }$ & $ 12.71 ^{+ 0.60 } _{- 0.53 }$
        & $ 17.01 ^{+ 0.28 } _{- 0.25 }$ & $15.64^{+0.09}_{-0.08}$
        \\ \hline
        
        NGC 6822 & $500\pm8$ & $18.19^{+0.10}_{-0.15}$ & $ 17.59 ^{+ 0.22 } _{- 0.31 }$ & $ 17.21 ^{+ 0.36 } _{- 0.40 }$
        & $ 18.64 ^{+ 0.08 } _{- 0.12 }$ & $17.30^{+0.04}_{-0.03}$
        \\ \hline
        
        PegDIG & $900\pm100$ & $16.73^{+0.11}_{-0.09}$ & $ 15.30 ^{+ 0.24 } _{- 0.18 }$ & $ 14.14 ^{+ 0.40 } _{- 0.29 }$ 
        & $ 17.67 ^{+ 0.17 } _{- 0.14 }$ & $16.52^{+0.04}_{-0.04}$
        \\ \hline
        
        Sculptor Dwarf Galaxy & $84\pm2$ & $18.23^{+0.04}_{-0.04}$ & $ 16.20 ^{+ 0.23 } _{- 0.22 }$ & $ 14.61 ^{+ 0.31 } _{- 0.33 }$ 
        & $ 18.21 ^{+ 0.06 } _{- 0.06 }$ & $17.42^{+0.02}_{-0.02}$
        \\ \hline
        
        Sextans dSph & $86\pm6$ & $19.42^{+0.09}_{-0.10}$ & $ 18.54 ^{+ 0.17 } _{- 0.17 }$ & $ 17.92 ^{+ 0.34 } _{- 0.26 }$
        & $ 18.99 ^{+ 0.06 } _{- 0.05 }$ & $18.48^{+0.04}_{-0.04}$
        \\ \hline
        
        Sgr dIG & $1040\pm50$ & $16.21^{+0.25}_{-0.17}$ & $ 14.49 ^{+ 0.56 } _{- 0.48 }$ & $ 12.99 ^{+ 0.80 } _{- 0.68 }$
        & $ 17.19 ^{+ 0.41 } _{- 0.40 }$ & $15.51^{+0.07}_{-0.05}$
        \\ \hline
        
        UMi Galaxy & $76\pm4$ & $19.13^{+0.20}_{-0.16}$ & $ 17.47 ^{+ 0.57 } _{- 0.70 }$ & $ 16.17 ^{+ 0.89 } _{- 1.06 }$
        & $ 18.64 ^{+ 0.18 } _{- 0.20 }$ & $17.61^{+0.10}_{-0.09}$
        \\ \hline
        
        WLM Galaxy & $920^{+43}_{-41}$ & $17.16^{+0.15}_{-0.13}$ & $ 15.96 ^{+ 0.34 } _{- 0.24 }$ & $ 15.02 ^{+ 0.47 } _{- 0.37 }$
        & $ 17.84 ^{+ 0.22 } _{- 0.23 }$ & $16.44^{+0.05}_{-0.05}$
        \\ \hline
        
        Z 64-73 (Leo I)& $300\pm50$& $17.98^{+0.08}_{-0.06}$ & $ 16.44 ^{+ 0.23 } _{- 0.27 }$ & $ 15.18 ^{+ 0.40 } _{- 0.34 }$
        & $ 18.09 ^{+ 0.15 } _{- 0.14 }$ & $17.01^{+0.03}_{-0.03}$
        \\ \hline
        
        Z 126-111 (Leo II)& $233\pm15$  & $17.52^{+0.18}_{-0.11}$ & $ 15.64 ^{+ 0.49 } _{- 0.45 }$ & $ 14.11 ^{+ 0.68 } _{- 0.65 }$
        & $17.88 ^{+ 0.21 } _{- 0.23 }$ & $16.52^{+0.06}_{-0.04}$
        \\ \hline
    \end{tabular}
    }
\caption{
\label{tab:results}
Calculated $J_s$-, $J_p$-, $J_d$- and $D$-factors for a set of dwarf spheroidal galaxies based on the estimates of the galactic DM profiles and velocity distributions presented in Ref.\,\cite{Bezrukov:2024qwr}. The error bars refer to $1\,\sigma$ confidence interval.
}
\end{table}

Astrophysical decay ($D$-factors) and $s$-annihilation factors ($J_s$-factors in our notations) for dwarf galaxies were estimated in a number of papers, see e.g.\,\cite{Geringer-Sameth:2014yza,Hayashi:2016kcy,Hayashi:2020jze,Yang:2025mye}. Results for individual galaxies generically vary form analysis to analysis but not much, typically within one or two standard deviations. Say, our results for Draco, Sextans and Ursa Minor deviate from the most recent results \cite{Yang:2025mye} by more than one error bar, our result for the Leo I are in agreement with those of Ref.\,\cite{Pascale:2025zga}  
while our error bars are much smaller. 
Since for any method used to reconstruct DM phase-space distribution the galaxy observables must be reproduced, it is more likely that the differences between our and previous works lay in the different stellar data rather than in different 
fitting formula for the DM galaxy profile $\rho_{DM}(r)$ and calculation methods. Say, the stellar samples we use for  Draco, Sextans and Ursa Minor are about twice of those used in Ref.\,\cite{Yang:2025mye}. Some analysis adopt the axisymmetric distributions for the stellar component, which generally increases the resulting uncertainties. The dominant parts of $D$- and $J$-factors are saturated within the galaxy half-light radius, where DM distribution is determined most accurately by analysis of the stellar data (see e.g. Fig. 1 in Ref.\,\cite{Bezrukov:2024qwr} and corresponding discussion there). This suggests that in the spherical DM assumption the choice of the functional form of $\rho_{DM}(r)$ made in this work does not lead to significant errors in determination of the $J$ and $D$ factors. 

It is worth noting that GravSphere does not account for the distance error when evaluating the scattering of the DM profiles. To verify that this inaccuracy leads to errors much smaller than the error bars of the geometric factors presented in Tab.\,\ref{tab:results}, we increased the distance to the Sculptor Dwarf Galaxy by 2\,kpc (that is about 1\,$\sigma$ error) and repeat the whole procedure. The decimal logarithm of 
$J$-factor obtained in this way is
 $18.25^{+0.05}_{-0.04}$, so its central value differs from the one given in Tab.\,\ref{tab:results} by less than 0.2\% and is well within 1\,$\sigma$ error bar. 

 On the other hand, the GravSphere solutions we use take into account the possible asymmetry in the DM velocities, which is the novelty of our analysis. One can expect this asymmetry affects the $J_p$- and $J_d$-factors given their explicit dependence on the asymmetry factor $\beta(r)$, see Eqs.\,\eqref{Jp}, \eqref{Jd}. This asymmetry cannot be directly extracted from observations and we adopted some priors to estimate its possible influence on the $J$-factors.

\begin{figure}[!htb]
    \centering
    \includegraphics[width=0.5\linewidth]{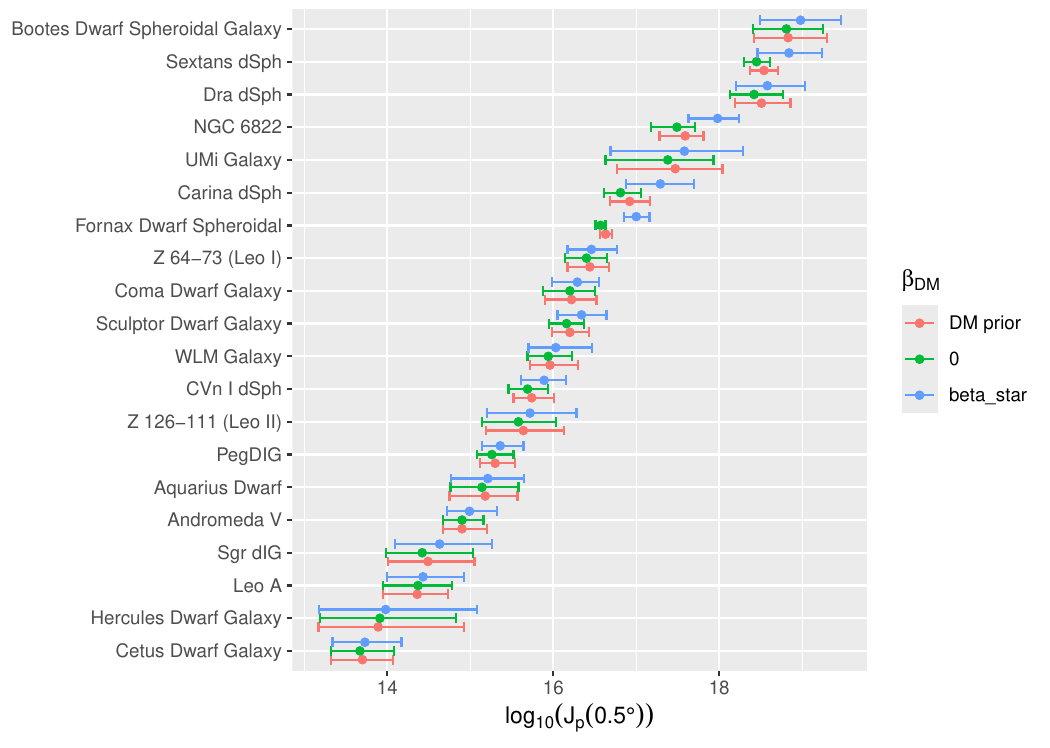}
    \includegraphics[width=0.5\linewidth]{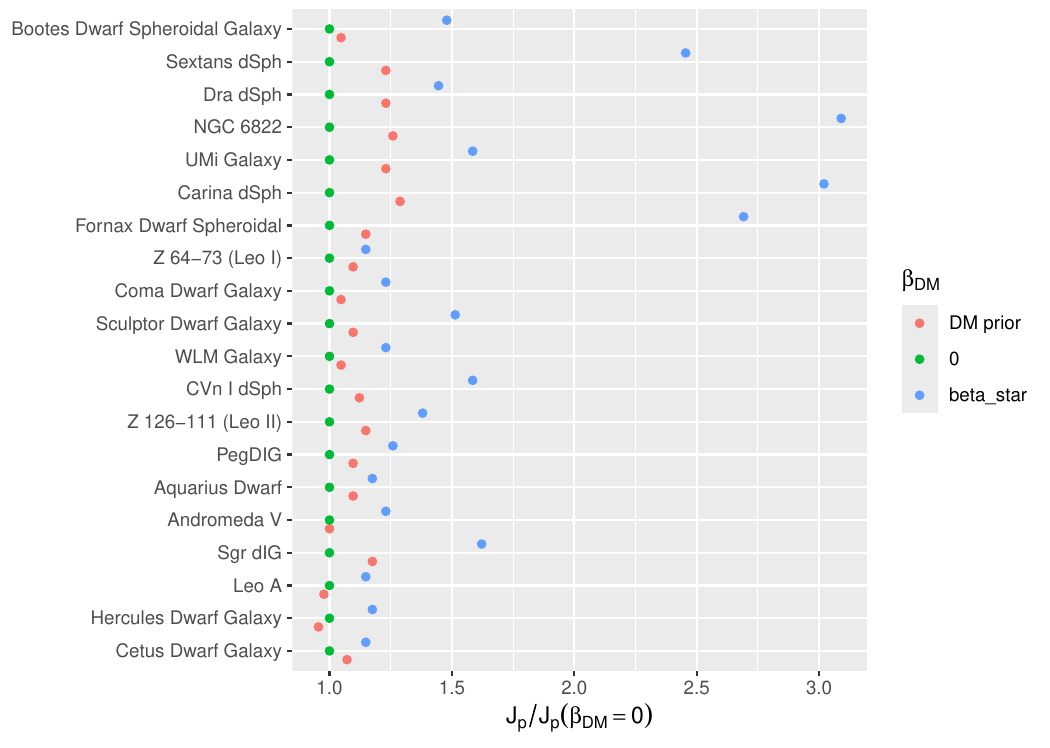}
    \caption{Comparison of $J_p$ calculation with different assumptions on $\beta_{DM}$.}
    \label{fig:beta_assumptions}
\end{figure}

To illustrate the model variability, we present in Tab.\,\ref{tab:var-asym}  
\begin{table}[!htb]
\centering    
\resizebox{\columnwidth}{!}{ 
    \begin{tabular}{|l|c|c|c|c|c|c|}
    
    \hline
        Name in SIMBAD database& $d/$\text{kpc}&
        $\log_{10}\frac{J_p(0.5^\circ,\beta_{DM}=0)}{ \text{GeV}^{2}\text{cm}^{-5}}$ & 
        $\log_{10}\frac{J_d(0.5^\circ,\beta_{DM}=0)}{\text{GeV}^{2}\text{cm}^{-5}}$ &
        $\log_{10}\frac{J_p(0.5^\circ, \beta_{DM}=\beta_*)}{ \text{GeV}^{2}\text{cm}^{-5}}$ & 
        $\log_{10}\frac{J_d(0.5^\circ\beta_{DM}=\beta_*)}{\text{GeV}^{2}\text{cm}^{-5}}$ \\ \hline  \hline
        
        Andromeda V & $810\pm45$ & $ 14.90 ^{+ 0.26 } _{- 0.23 }$ & $ 13.41 ^{+ 0.39 } _{- 0.32 }$
        & $ 14.99 ^{+ 0.33 } _{- 0.27 }$ & $ 13.58 ^{+ 0.54 } _{- 0.43 }$
        \\ \hline
        
        Aquarius Dwarf & $940\pm38$ & $ 15.14 ^{+ 0.44 } _{- 0.38 }$ & $ 13.78 ^{+ 0.52 } _{- 0.56 }$
        & $ 15.21 ^{+ 0.44 } _{- 0.44 }$ & $ 13.90 ^{+ 0.68 } _{- 0.64 }$
        \\ \hline
        
        Bootes Dwarf Spheroidal Galaxy & $66\pm3$  & $ 18.81 ^{+ 0.44 } _{- 0.40 }$ & $ 17.99 ^{+ 0.60 } _{- 0.57 }$
        & $ 18.98 ^{+ 0.49 } _{- 0.49 }$ & $ 18.34 ^{+ 0.81 } _{- 0.67 }$
        \\ \hline
        
        Carina dSph & $105\pm6$  & $ 16.81 ^{+ 0.25 } _{- 0.20 }$ & $ 15.58 ^{+ 0.38 } _{- 0.34 }$
        & $ 17.29 ^{+ 0.41 } _{- 0.41 }$ & $ 16.62 ^{+ 0.91 } _{- 0.77 }$
        \\ \hline
        
        Cetus Dwarf Galaxy & $790\pm40$  & $ 13.67 ^{+ 0.41 } _{- 0.35 }$ & $ 11.78 ^{+ 0.49 } _{- 0.46 }$
        & $ 13.73 ^{+ 0.44 } _{- 0.39 }$ & $ 11.89 ^{+ 0.67 } _{- 0.48 }$
        \\ \hline
        
        Coma Dwarf Galaxy & $44\pm4$  & $ 16.20 ^{+ 0.30 } _{- 0.32 }$ & $ 14.41 ^{+ 0.41 } _{- 0.41 }$
        & $ 16.29 ^{+ 0.26 } _{- 0.31 }$ & $ 14.74 ^{+ 0.34 } _{- 0.42 }$
        \\ \hline
        
        CVn I dSph & $218\pm10$  & $ 15.69 ^{+ 0.25 } _{- 0.23 }$ & $ 14.19 ^{+ 0.36 } _{- 0.32 }$
        & $ 15.89 ^{+ 0.26 } _{- 0.28 }$ & $ 14.53 ^{+ 0.62 } _{- 0.44 }$
        \\ \hline
        
        Dra dSph & $76\pm5$  & $ 18.42 ^{+ 0.35 } _{- 0.29 }$ & $ 17.58 ^{+ 0.48 } _{- 0.42 }$
        & $ 18.58 ^{+ 0.45 } _{- 0.38 }$ & $ 17.91 ^{+ 0.87 } _{- 0.68 }$
        \\ \hline
        
        Fornax Dwarf Spheroidal & $147\pm12$ & $ 16.57 ^{+ 0.06 } _{- 0.06 }$ & $ 15.28 ^{+ 0.09 } _{- 0.10 }$
        & $ 17.00 ^{+ 0.16 } _{- 0.15 }$ & $ 16.16 ^{+ 0.38 } _{- 0.40 }$
        \\ \hline
        
        Hercules Dwarf Galaxy & $132\pm6$  & $ 13.91 ^{+ 0.92 } _{- 0.72 }$ & $ 11.69 ^{+ 1.22 } _{- 1.08 }$
        & $ 13.98 ^{+ 1.10 } _{- 0.80 }$ & $ 11.77 ^{+ 1.38 } _{- 1.05 }$
        \\ \hline
        
        Leo A & $800\pm100$  & $ 14.37 ^{+ 0.41 } _{- 0.42 }$ & $ 12.67 ^{+ 0.57 } _{- 0.49 }$
        & $ 14.43 ^{+ 0.49 } _{- 0.43 }$ & $ 12.96 ^{+ 0.86 } _{- 0.66 }$
        \\ \hline
        
        NGC 6822 & $500\pm8$  & $ 17.49 ^{+ 0.22 } _{- 0.31 }$ & $ 17.05 ^{+ 0.32 } _{- 0.46 }$
        & $ 17.98 ^{+ 0.26 } _{- 0.35 }$ & $ 18.37 ^{+ 0.53 } _{- 0.69 }$
        \\ \hline
        
        PegDIG & $900\pm100$  & $ 15.26 ^{+ 0.26 } _{- 0.18 }$ & $ 14.05 ^{+ 0.35 } _{- 0.29 }$
        & $ 15.36 ^{+ 0.28 } _{- 0.22 }$ & $ 14.18 ^{+ 0.57 } _{- 0.38 }$
        \\ \hline
        
        Sculptor Dwarf Galaxy & $84\pm2$  & $ 16.16 ^{+ 0.21 } _{- 0.21 }$ & $ 14.55 ^{+ 0.30 } _{- 0.32 }$
        & $ 16.34 ^{+ 0.30 } _{- 0.29 }$ & $ 14.92 ^{+ 0.59 } _{- 0.50 }$
        \\ \hline
        
        Sextans dSph & $86\pm6$  & $ 18.45 ^{+ 0.16 } _{- 0.15 }$ & $ 17.75 ^{+ 0.27 } _{- 0.21 }$
        & $ 18.84 ^{+ 0.40 } _{- 0.38 }$ & $ 18.44 ^{+ 0.87 } _{- 0.69 }$
        \\ \hline
        
        Sgr dIG & $1040\pm50$  & $ 14.42 ^{+ 0.61 } _{- 0.44 }$ & $ 12.93 ^{+ 0.83 } _{- 0.69 }$
        & $ 14.63 ^{+ 0.63 } _{- 0.54 }$ & $ 13.37 ^{+ 1.00 } _{- 0.92 }$
        \\ \hline
        
        UMi Galaxy & $76\pm4$  & $ 17.38 ^{+ 0.55 } _{- 0.75 }$ & $ 16.06 ^{+ 0.88 } _{- 1.03 } $
        & $ 17.58 ^{+ 0.71 } _{- 0.89 }$ & $ 16.33 ^{+ 1.18 } _{- 1.17 }$
        \\ \hline
        
        WLM Galaxy & $920^{+43}_{-41}$  & $ 15.94 ^{+ 0.29 } _{- 0.25 }$ & $ 14.94 ^{+ 0.42 } _{- 0.37 }$
        & $ 16.03 ^{+ 0.44 } _{- 0.33 }$ & $ 15.21 ^{+ 0.81 } _{- 0.53 }$
        \\ \hline
        
        Z 64-73 (Leo I)& $300\pm50$  & $ 16.40 ^{+ 0.25 } _{- 0.26 }$ & $ 15.13 ^{+ 0.33 } _{- 0.34 }$
        & $ 16.46 ^{+ 0.31 } _{- 0.29 }$ & $ 15.29 ^{+ 0.59 } _{- 0.48 }$
        \\ \hline
        
        Z 126-111 (Leo II)& $233\pm15$  & $ 15.58 ^{+ 0.45 } _{- 0.44 }$ & $ 14.03 ^{+ 0.69 } _{- 0.61 }$
        & $ 15.72 ^{+ 0.56 } _{- 0.52 }$ & $ 14.3 ^{+ 0.90 } _{- 0.81 }$
        \\ \hline
    \end{tabular}
    }
\caption{The values of $J_p$- and $J_d$-factors inferred for different assumptions about the DM velocity asymmetry $\beta_{DM}$.
}
\label{tab:var-asym}
\end{table}
the estimates of $J_p$- and $J_d$-factors obtained with 1) purely isotropic DM velocities, 2) DM velocities with the same asymmetry as observed in stellar population of each particular galaxy. One finds that in the first case the values of $J$-factors are systematically smaller, while in the second case they are systematically larger, than those with DM priors. However, the central values always stay well within the 1\,$\sigma$ error bars. We illustrate the distributions of the estimates of individual $J_p$-factors on the plots of Fig.\,\ref{fig:beta_assumptions}.
Note, that 
in the case of the DM following the star velocity asymmetry the values obtained for four galaxies, namely Sextans, NGC 6822, Carina and Fornax, deviate significantly (by a factor 2.5-3) from those obtained assuming symmetric DM velocities, which may be treated as the most conservative case providing the smallest values of $J$-factors. We discuss this issue in due course.

{\bf 4.} Having obtained the geometric factors for a set of galaxies we can find the scaling relations which allow to get the reliable estimates of the factors using only the direct galaxy observables: the line-of-sight stellar velocity dispersion $\sigma_{\text{LOS}}$, the distance to the galaxy $d$ and the half-light radius $r_h$. Such power-law relations for $D$-factors and $J_s$-factors  have been suggested in Ref.\,\cite{Pace:2018tin}. Here we refine them, and further suggest similar scaling relations for $J_p$- and $J_d$-factors. 

The power-law relations introduced in Ref.\cite{Pace:2018tin} for $D$- and $J_s$-factors are parametrically the same, 
\begin{align}
    \label{D-scaling}
D & = D_0 \l\frac{\sigma_{LOS}}{5\, \text{km/s}}\r^{\gamma_{\sigma_{LOS}}} \l\frac{d}{100\,\text{kpc}}\r^{\gamma_{d}} \l\frac{r_h}{100 \text{\;pc}}\r^{\gamma_{r_h}}\,,\\
J_{s} & = J_{s0} \l\frac{\sigma_{LOS}}{5\, \text{km/s}}\r^{\gamma_{\sigma_{LOS}}} \l\frac{d}{100\,\text{kpc}}\r^{\gamma_{d}} \l\frac{r_h}{100 \text{\;pc}}\r^{\gamma_{r_h}}\,.
    \label{Js-scaling}
\end{align}
With this power-low form we fit our results for the $D$-factors presented in Tab.\,\ref{tab:results}. The obtained fitting parameters are shown in Tab.\,\ref{tab:D-factor} 
\begin{table}[!htb]
    \centering
    \begin{tabular}{|c|c|c|c|c|}
    \hline
     &$\log_{10}\l \frac{D_0}{\text{GeV}^{-1}\text{cm}^2} \r$ & $\gamma_{\sigma_{LOS}}$ & $\gamma_{d}$ & $\gamma_{r_{h}}$ \\ \hline\hline
     Our fit & $17.70\pm 0.12$ & $2.33\pm 0.27$ & $-1.17\pm 0.13$ & $0.08\pm 0.27$ \\ \hline
     fit of Ref.\,\cite{Pace:2018tin} & 1$7.93^{+0.09}_{-0.10}$ & $1.7\pm 0.5$ & $-0.9\pm0.2$ & $-0.5\pm0.3$ \\ \hline \hline
     Our fit & $17.53\pm 0.22$ & $1.94 \pm 0.47 $& -2 & $1.27\pm0.34$ \\ \hline
      fit of Ref.\,\cite{Pace:2018tin} & - & $\approx 3.4$ & -2 & $\approx -0.05$ \\ \hline \hline
      Our fit $\theta_{max}=\alpha_c/2$ & $16.64\pm0.03$ & $1.84\pm0.06$ & $-2.01\pm0.03$ & $1.08\pm0.06$ \\ \hline
      fit of Ref.\,\cite{Pace:2018tin} $\theta_{max}=\alpha_c/2$ & $16.65\pm 0.04$ & $1.9\pm0.2$ & $-1.9\pm0.6$ & $0.9\pm0.1$ \\ \hline
      \hline
      Our fit $\theta_{max}=\alpha_c/2$ & $16.63\pm0.01$ & 2 & -2 & 1 \\ \hline
      fit of Ref.\,\cite{Pace:2018tin} $\theta_{max}=\alpha_c/2$ & $16.57\pm0.02$ & 2 & -2 & 1 \\ \hline
    \end{tabular}
    \caption{
    \label{tab:D-factor}
    The results of our fits of $D$-factor and its comparison with those in literature.}
\end{table}
along with the fitting parameters of Ref.\,\cite{Pace:2018tin}. Our results deviate from those of Ref.\,\cite{Pace:2018tin} within 1-2 standard deviations. Since the signal of the same galaxy would scale as the inverse squared distance, one can fix $\gamma_d=-2$ and fit the $D$-factors of the 20 galaxies with only three parameters. Our results deviate significantly from that of 
Ref.\,\cite{Pace:2018tin}, though it does not contain any estimates of the errors for that fit.  Once $\gamma_d$ is fixed, the normalization becomes a little smaller, well within the error bars.

Other lines contained the parameters of the fit obtained when for each galaxy the integration over the angle is performed in the range $\theta\in[0;\theta_{max}]$, where the maximal angle is chosen individually as $\theta_{max}=\frac{\alpha_c}{2}\approx\frac{r_h}{d}$. Then from the equation for the tangential velocity dispersion (2nd Newtonian law) one expects the simple scaling relation between the total galaxy mass contained within the sphere of radius $r_h$ and the two fitting parameters, $M(r_h)\propto \sigma^2 r_h$. The obtained numbers confirm this expectation. We also present the results of the fit where the only free parameter is the normalization factor, once all the exponents are fixed as explained above. 
With the maximal integration angle chosen individually, the error bars in the $J$-factors become smaller, while the normalization factor remains almost intact. However, it deviates significantly from that obtained with integration over $0.5^\circ$.

\begin{figure}[!htb]
    \centerline{
    \includegraphics[width=\columnwidth/2]{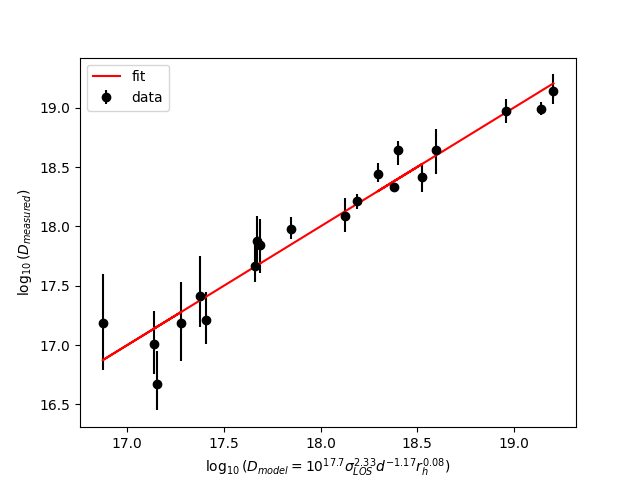}
    \includegraphics[width=\columnwidth/2]{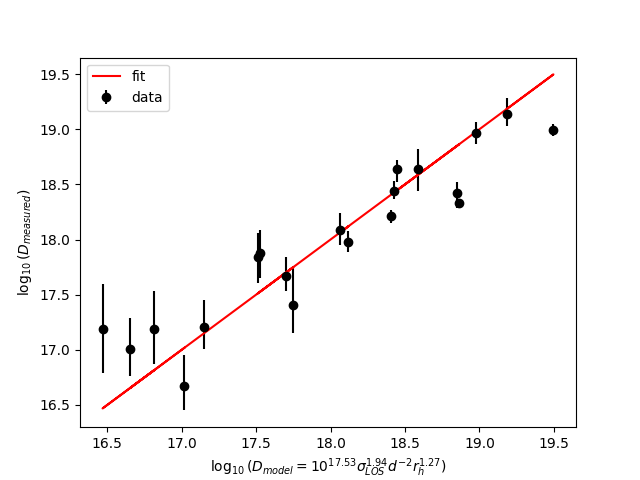}
    }
    \caption{
    \label{fig:D}
    Distributions of the 20 dSphs over the fitting parameters for the two variants of the fit\,\eqref{D-scaling}: $\gamma_d$ is free (left panel) and $\gamma_d=-2$ (right panel).}
\end{figure}

Two our variants of the fit, that with free parameters $\gamma_d$ and that with $\gamma_d=-2$, are illustrated with distribution of the all 20 galaxies in Fig.\,\ref{fig:D}. 

\begin{figure}[!htb]
    \centerline{
    \includegraphics[width=\columnwidth/2]{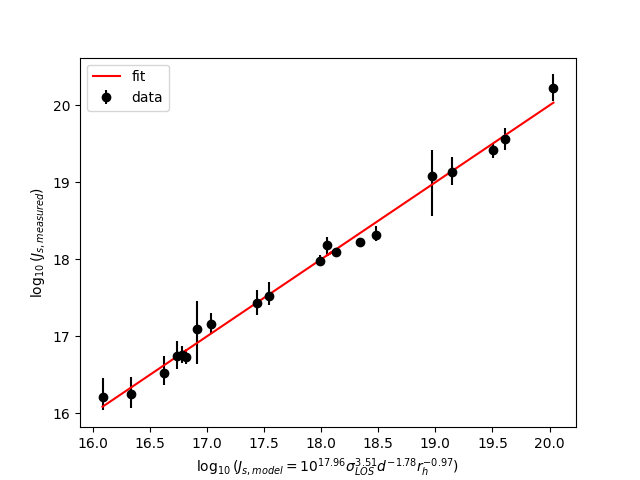}
    \includegraphics[width=\columnwidth/2]{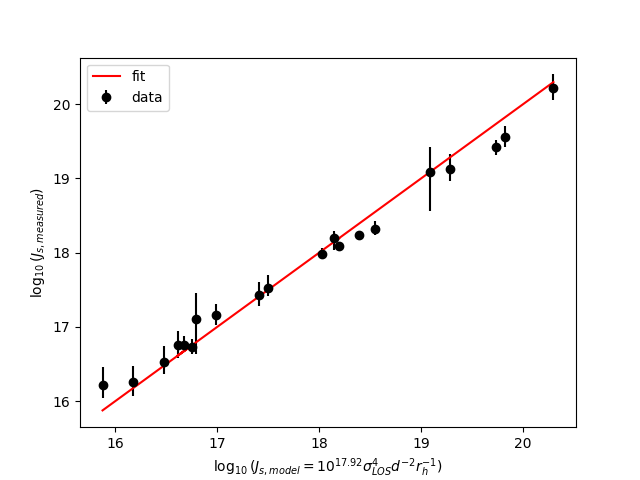}
    }
    \caption{
    \label{fig:Js}
    The two versions of our fit to $J_s$-factor obtained from analysis\,\cite{Bezrukov:2024qwr} of the stellar observations of 20 dSphs.}
\end{figure}

Similarly, we fit the $J_s$-factors with the scaling relation,\eqref{Js-scaling}, our results and those of Ref.,\cite{Pace:2018tin} are presented in Tab.,\ref{tab:Js-factor}. 
\begin{table}[!htb]
    \centering
    \begin{tabular}{|c|c|c|c|c|}
    \hline
     & $\log_{10}\frac{J_0}{\text{GeV}^2\text{cm}^5}$ & $\gamma_{\sigma_{LOS}}$ & $\gamma_{d}$ & $\gamma_{r_{h}}$ \\ \hline \hline 
     Our fit & $17.96\pm 0.07$ & $3.51\pm 0.15$ & $-1.78\pm 0.08$ & $-0.97\pm 0.15$ \\ \hline
      fit of Ref.\,\cite{Pace:2018tin}  & $17.96\pm 0.9$ & $3.8\pm 0.4$ & $-1.8\pm 0.1$ & $-1.2\pm 0.2$ \\ \hline \hline
     Our fit & $17.92\pm 0.04$ & 4 & -2 & -1 \\ \hline
      fit of Ref.\,\cite{Pace:2018tin}  & $17.87^{+0.04}_{-0.03}$ & 4 & -2 & -1 \\ \hline
    \end{tabular}
    \caption{
    \label{tab:Js-factor}
    Comparison between our best-fit parameters for $J_s$-factors and ones suggested in literature.}
\end{table}
The two variants of the fit, with free $\gamma_d$ and $\gamma_d=-2$ are presented. In the last case, the exponents of half-light radius and line-of-sight dispersions have also been fixed, motivated by the scaling we discussed above, $M(r_h)\propto \sigma^2 r_h$, and the order-of-magnitude estimate of the DM density $\rho\propto M(r_h)/r_h^3$. Together, they provide with the scaling $J_s\propto \sigma^4/r_h$. 
The two our fits are illustrated in Fig.\,\ref{fig:Js}.

The scaling relations illustrate the suggested explanation of the deviations between the different analyses for the individual galaxies. While with different stellar samples the observable parameters like $\sigma_{los}$, $d$, $r_h$ are different, the values of $D$- and $J$-factors estimated for the individual galaxies with the general scaling relations, obtained in our paper, and those, obtained in Refs.\,\cite{McDaniel:2023bju,Yang:2025mye}, coincide within 1\,$\sigma$.    

\begin{figure}[!htb]
    \centerline{
    \includegraphics[width=\columnwidth/2]{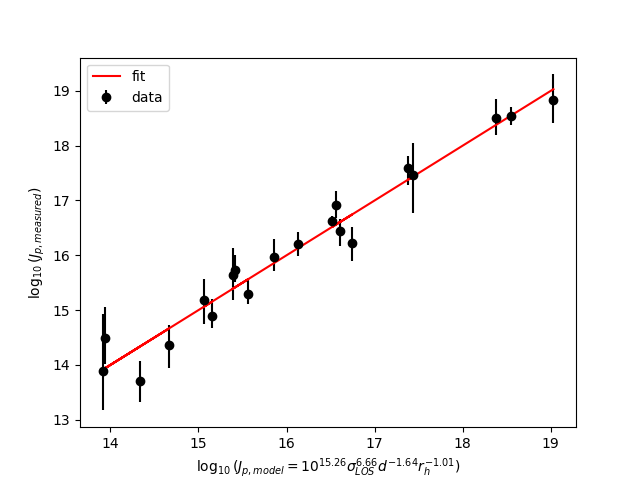}
    \includegraphics[width=\columnwidth/2]{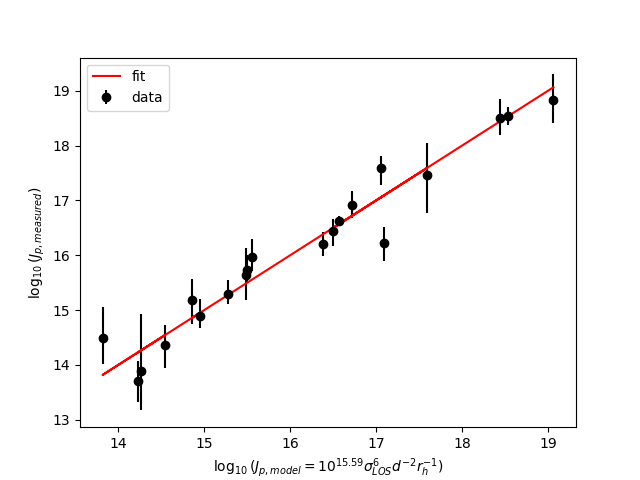}
    }
    \centerline{
    \includegraphics[width=\columnwidth/2]{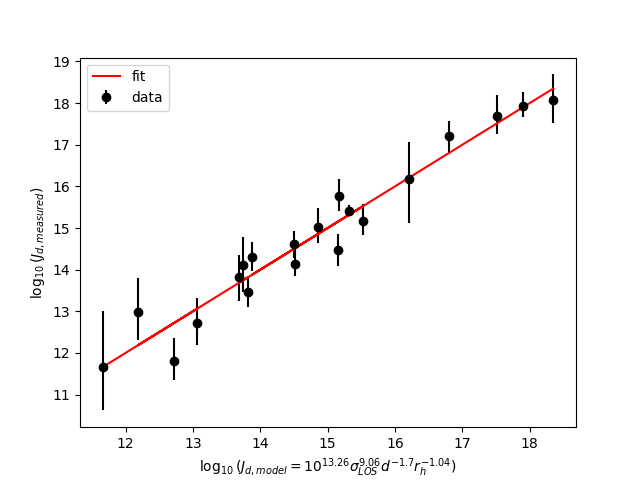}
    \includegraphics[width=\columnwidth/2]{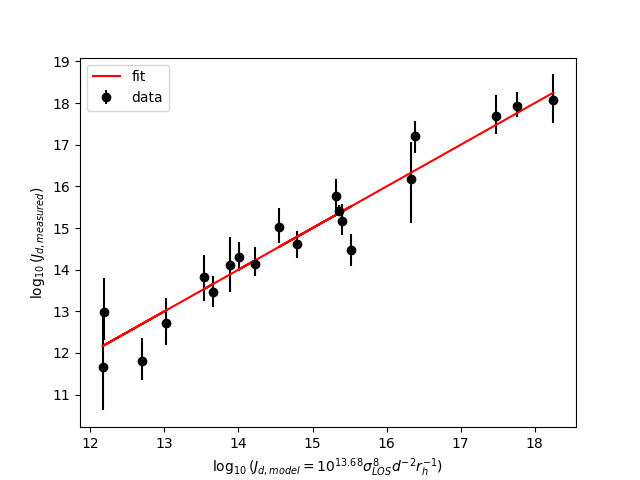}
    }
    \caption{
    \label{fig:JpJd-factors}
    Our fits to $J_p$- (top panels) and $J_d$-factors (bottom panels).}
\end{figure}

Finally, we perform similar fits for the $J_p$- and $J_d$-factors. The simple considerations suggest the scaling relations $J_p\propto \sigma^6/r_h$, $J_d\propto \sigma^8/r_h$, since the integrals \eqref{Jp}, \eqref{Jd} contain the corresponding velocity-dependent factors. Results of our fits are shown in Tab.\,\ref{tab:JpJd-factors}. 
\begin{table}[!htb]
    \centering
    \begin{tabular}{|c|c|c|c|}
    \hline
     $\log_{10}\frac{J_{p,0}}{\text{GeV}^2\text{cm}^5}$ & $\gamma_{p,\sigma_{LOS}}$ & $\gamma_{p,d}$ & $\gamma_{p,r_{h}}$ \\ \hline
     $15.26\pm 0.21$ & $6.66\pm 0.44$ & $-1.64\pm 0.22$ & $-1.01\pm 0.44$ \\ \hline
     $15.59\pm 0.08$ & 6 & -2 & -1 \\ \hline \hline 
     $\log_{10}\frac{J_{d,0}}{\text{GeV}^2\text{cm}^5}$  & $\gamma_{d,\sigma_{LOS}}$ & $\gamma_{d,d}$ & $\gamma_{d,r_{h}}$ \\ \hline
     $13.26\pm 0.30$ & $9.06\pm 0.64$ & $-1.70\pm 0.32$ & $-1.04\pm 0.64$ \\ \hline
     $13.68\pm 0.11$ & 8 & -2 & -1 \\ \hline
    \end{tabular}
    \caption{
    \label{tab:JpJd-factors}
    Our fits for $J_p$- and $J_d$-factors.}
\end{table}
The fits are illustrated with plots in Fig.\,\ref{fig:JpJd-factors}.

{\bf 5.} To conclude, let us summarise the obtained results. 

While the observations of the stellar dynamics gives insight into the gravitational potential of a given galaxy, and, thus, into the distribution of DM in this galaxy, the velocities of the DM escape even indirect measurements. Note, that even stellar anisotropy is very hard to estimate form current observations. Really, the Jeans equation relating DM density profile and velocity dispersions depend on the DM velocity anisotropy. If DM annihilates in p- or d-channel, this velocity ambiguity also enters into the potential annihilation signal.

While we can not measure the DM velocity anisotropy, we can assume it to be either absent (what was done in all previous estimates of $J$ factors for its annihilation), make an educated guess about its value based on DM simulations, or even assume it to be similar to the stellar anisotropy. With increased anisotropy the DM velocities and respective $J_p$ factors increase. Our analysis shows, that our ignorance of $\beta_{DM}$ really introduces change in $J_p$ factor, but for the majority of galaxies the effect, though present, is well within the uncertainties of the observations (c.f.~Fig.\ref{fig:beta_assumptions}).

We updated the estimates for decay and s-wave annihilation $D$ and $J_s$ factors based on the latest dwarf spheroidal observations, including only sufficiently large galaxies allowing for reliable DM fits. We also repeated the empiric fits to obtain scaling relations for the factors depending on overall galaxy parameters: its distance, line of sight velocity dispersion and half-light radius. The largest observe difference was found in for the $D$ factor scaling relation.

In addition, we obtained $J_p$ and $J_d$ factors, corresponding to the models where s-wave annihilation of the DM is impossible, and derived simple scaling relations for them too.  The results allow for obtaining new bounds on DM models from the annihilation signal from DM dominated galaxies. 
 These factors explicitly depend on the DM velocity distribution, including the asymmetry. The factors for individual galaxies change with the assumption about the asymmetry, but well within the error bars. The scaling relations we obtained for the whole galaxy set remain almost intact with respect to change of the assumptions, see Tab.\,\ref{tab:as}. 
\begin{table}[!htb]
    \centering{
    \begin{tabular}{|c|c|c|c|c|}
    \hline
     & $\log_{10}\frac{J_{p,0}}{\text{GeV}^2\text{cm}^5}$ & $\gamma_{\sigma_{LOS}}$ & $\gamma_{d}$ & $\gamma_{r_{h}}$ \\ \hline \hline 
     $\beta_{DM}=0$ & $15.23\pm 0.19$ & $6.57\pm 0.41$ & $-1.62\pm 0.20$ & $-1.01\pm 0.41$ \\ \hline
     $\beta_{DM}=0$ & $15.55\pm0.08$ & 6 & -2 & -1 \\ \hline
      $\beta_{DM}=\beta_*$  & $15.27\pm 0.25$ & $6.80\pm 0.54$ & $-1.81\pm 0.26$ & $-0.72\pm 0.54$ \\ \hline
      $\beta_{DM}=\beta_*$ & $15.73\pm0.09$ & 6 & -2 & -1 \\ \hline \hline
      & $\log_{10}\frac{J_{d,0}}{\text{GeV}^2\text{cm}^5}$ & $\gamma_{\sigma_{LOS}}$ & $\gamma_{d}$ & $\gamma_{r_{h}}$ \\ \hline \hline
     $\beta_{DM}=0$ & $13.23\pm 0.28$ & $8.89\pm0.61$ & $-1.65\pm0.30$ & $-1.10\pm0.61$ \\ \hline
     $\beta_{DM}=0$ & $13.60\pm0.10$ & 8 & -2 & -1 \\ \hline
      $\beta_{DM}=\beta_*$  & $13.28\pm0.42$ & $9.49\pm0.91$ & $-1.98\pm0.45$ & $-0.53\pm0.91$ \\ \hline
      $\beta_{DM}=\beta_*$ & $13.99\pm0.15$ & 8 & -2 & -1 \\ \hline
    \end{tabular} 
    }
    \caption{ Our fits for $J_p$- and $J_d$- factors for different choices of $\beta_{DM}$.}
    \label{tab:as}
\end{table}
Recall that four galaxies in the sample, namely Sextans, NGC 6822, Carina and Fornax, exhibit noticeable change in the individual values of the velocity-dependent $J_p$- and $J_d$-factors illustrated with plots of Fig.\,\ref{fig:beta_assumptions}. We checked that their exclusion from the galaxy sample shifts the parameters of the scaling relations well within the error bars.

\vskip 0.3cm

The research is partially supported by the National Center of Physics and Mathematics, direction No.\,5. 
EK thanks the Foundation for the Advancement of Theoretical Physics and Mathematics “BASIS” for the individual grant.


\vskip 1cm

\bibliographystyle{utphys}
\bibliography{refs}
\end{document}